\documentclass{ws-p9-75x6-50}

\def\sqr#1#2{{\vcenter{\hrule height.#2pt \hbox{\vrule width.#2pt
height#1pt \kern#1pt \vrule width.#2pt} \hrule height.#2pt}}}

\def \sc{\scriptstyle}
\def \darrow#1{\raise1.5ex\hbox{$\rightarrow$}\mkern-16.5mu #1}
\def\darr#1{\raise1.5ex\hbox{$\leftrightarrow$}\mkern-16.5mu #1}

\def\rpartial{\darrow{\partial}}

\def\phi{\varphi}

\begin{document}

\title{Scheme independence as an inherent redundancy in quantum 
field theory}

\author{Jos\'e I. Latorre}

\address{Departament d'Estructura i Constituents de la Mat\`eria,
Universitat de Barcelona, and I.F.A.E.,
Diagonal 647, E-08028 Barcelona, Spain}

\author{Tim R. Morris}

\address{Department of Physics, University of Southampton,
Highfield, Southampton SO17~1BJ, UK
%\\E-mail: T.R.Morris@soton.ac.uk
}

\maketitle

\abstracts{
The path integral formulation of Quantum Field Theory implies
an infinite set of local, Schwinger-Dyson-like relations.
Exact renormalization
group equations can be cast as a particular instance of
these relations. Furthermore, exact scheme independence is 
turned into
a vector field transformation of the kernel of the exact
renormalization group equation under field redefinitions.
}

Relations for the  Green functions 
of a quantum field theory are in general
traced down to symmetries of the partition function. In particular,
the path integral allows for the basic symmetry of field redefinitions.
A vast family of Schwinger-Dyson like 
relations follows by taking the infinitessimal 
field redefinition
$\tilde\phi^\alpha=\phi^\alpha-\theta^\alpha[\phi,t]$
\be
\int {\cal D}\phi\
\partial_\alpha\left(\theta^\alpha {\rm e}^{-S} \right)=0\quad
\Rightarrow 
\langle \partial_\alpha \theta^\alpha \rangle =
 \langle \theta^\alpha\partial_\alpha S\rangle \ ,  
\label{fieldredefid}
\ee
where we have introduced the compact notation
\be
\label{notation}
  \partial_\alpha \equiv {\delta\over \delta \phi^\alpha} \ ,
\ee
$\alpha$ standing for momentum and other possible quantum numbers.

\bigskip

There is a deep connection between this observation and the 
exact renormalization group.\cite{red}
Remarkably, the exact renormalization group
equation itself,\cite{kogwil,wegho} 
as formulated by Polchinski,\cite{pol} can be written as
the integrand of a field redefinition transformation.\cite{alg,latmor}
To see this, let us first recall the dimensionless form of 
Polchinski's equation
\bea
\label{polchinski}
\partial_tS+d\int_p\!\phi_p{\delta S\over\delta\phi_p}
&+&\int_p\!\phi_p\,
  p^\mu{\partial\over\partial p^\mu}{\delta S\over\delta\phi_p}
= \nonumber\\
&{}&\phantom{\int_p\!\phi_p\,}
\int_p c'(p^2)
     \left( {\delta S\over \delta \phi_p}{\delta S\over \delta \phi_{-p}}
     - {\delta^2 S\over \delta \phi_p \delta \phi_{-p}}
  -2{p^2\over c(p^2)}\phi_p {\delta S\over \delta \phi_p}\right) \ ,
\eea
where $t\equiv \ln {\mu\over\Lambda}$, $\mu$ being some fixed 
physical scale
and $\Lambda$ the Wilsonian cutoff which is taken towards 0,
$S[\phi,t]$ is a functional of $\phi$ and a function of $t$ and
$c(p^2)$ is the regulating function defining classes of schemes.

We can now check that Polchinski's equation is identical to 
\be
\label{wegner}
  \partial_t\, {\rm e}^{-S}=
  \partial_\alpha\left( \Psi^\alpha {\rm e}^{-S}\right)\ ,
\ee
where the functional $\Psi^\alpha$ corresponds to
\be
\label{psidef}
\Psi^\alpha[\phi,t]= \left( D-d(t)\right)
\phi_p+p^\mu{\partial\phi_p\over\partial p^\mu}\,+\,
c'(p^2)\left({\delta S\over \delta \phi_{-p}}
     -2{p^2\over c(p^2)} \phi_p\right)\ .
\ee
Two facts are now clear:
\par$\bullet$ The equation is traced to the field redefinition
\be
\label{psiredef}
\tilde\phi^\alpha=\phi^\alpha-\delta
t\,\Psi^\alpha[\phi,t]\ .
\ee
\par$\bullet$
Upon integration, Polchinski's equation implies
\be
\label{newwegner}
 0=\partial_t \int {\cal D} \phi \ {\rm e}^{-S}
= \int {\cal D}\phi \  \partial_\alpha\left(\Psi^\alpha {\rm e}^{-S}
  \right)=0\ ,
\ee
that is, the l.h.s. vanishes because low-energy  
physics does not depend 
on the cut-off, whereas the r.h.s. is zero because
it is the total derivative term emerging from
a field redefinition. 

\bigskip

The field redefinition symmetry underlying the renormalization
group does also control its scheme dependence. It is easy
to compute the change of the kernel in the exact renormalization
group equation under a field redefinition. Starting from
$\Psi^\alpha$ and redefining $\phi^\alpha\rightarrow
\phi^\alpha-\theta^\alpha$, produces the change
\bea
\label{psitranf}
   \delta \Psi^\alpha[\phi,t]
       &\equiv& \hat\Psi^{\alpha}[\phi,t]-\Psi^\alpha[\phi,t] \nonumber\\
       &=& \partial_t \theta^\alpha
        +\theta^\beta\partial_\beta\Psi^\alpha-\Psi^\beta\partial_\beta
         \theta^\alpha\ .
\eea
This can  be further framed as a standard vector field transformation
plus the explicit $t$ derivative,
\be
\label{vectortransf}
\delta\Psi^\alpha \rpartial_\alpha =
 \partial_t \theta^\alpha  \rpartial_\alpha+
  \left[ \theta^\beta\rpartial_\beta,\Psi^\alpha\rpartial_\alpha\right]
\ .
\ee

\bigskip

The above equations  suggest a more compact notation. 
Let us 
put the differential on the left and define the following shorthand
\be
\label{gadf}
\theta\equiv \rpartial_\alpha\theta^\alpha\quad,\quad
A_t\equiv \rpartial_\alpha\Psi^\alpha\quad,\quad
\delta A_t\equiv \rpartial_\alpha\delta \Psi^\alpha\quad{\rm and}\quad
D_t\equiv\partial_t-A_t\ .
\ee
In this notation the exact RG (\ref{wegner}) simply reads
\be
\label{simplerg}
   D_t {\,\rm e}^{-S}=0\ .
\ee
This new form of the exact renormalization group equation 
deserves a more detailed inspection. Let us in turn
discuss its transformation properties and, then, its geometrical
interpretation.

By (\ref{fieldredefid}), 
the change in the action under a field redefinition is just
\be
\label{gas}
\delta {\,\rm e}^{-S}= \theta {\,\rm e}^{-S}
\ee
and, either by integrating by parts the derivative in (\ref{vectortransf}),
or directly from (\ref{psitranf}),
\be
\label{gaa}
\delta A_t=\partial_t\theta-[A_t,\theta]\ 
\equiv [D_t,\theta]\ .
\ee
Now it is easy to see that our transformation for $\Psi^\alpha$
does indeed ensure that the exact RG transforms covariantly:
\bea
\label{gaerg}
\delta\left(D_t {\,\rm e}^{-S}\right)
&=&-\delta A_t{\,\rm e}^{-S}+D_t\,\delta{\,\rm e}^{-S} \nonumber\\
&=&-[D_t,\theta]{\,\rm e}^{-S}+ D_t\, \theta{\,\rm e}^{-S} \nonumber\\
&=&\theta\left(D_t {\,\rm e}^{-S}\right)\ .
\eea

The functional $A_t$ emerges as a connection in the space of actions
fibered over renormalization group time. Local field redefinition
symmetry is maintained through the presence of $A_t$, which is
in charge of notifying to neighbouring points what choice of fields 
has been made. In this fashion the relative weight among field
configurations is modified along the flow. 

It is  natural to ask what are the (observable)
invariants associated to the general field redefinition 
symmetry. These may be local or global.
The fact that the renormalization group time is
one-dimensional explicitly forbids the existence of the former
type: there is no field strength in one dimension.
Only integrated invariants can be obtained. Indeed, the 
Wilson line starting from $t=t_0$ is of interest:
\be
\label{Ph}
\Phi(t)=P\exp\left(-\int^t_{t_0}\!\!\!\!ds\, A_s\right),
\ee
where $P$ is path ordering, placing as usual the later $A$s 
to the right (thus $\partial_t\Phi(t)=-\Phi(t) A_t$ and $\partial_t\Phi^{-1}(t)
=-A_t\Phi^{-1}(t)$).

Let us put together our line of thought. A particular
exact RG equation is characterized by its kernel $\Psi^\alpha$.
The freedom to choose the form of this kernel is  related to field
redefinitions. Infinitessimally, two kernels, expressed in terms of the
action $S$, are physically equivalent if they are related by 
\be
\label{fsda}
\delta \bar A_t[\phi,S]\equiv 
\tilde{\bar A}_t[\phi,S]-\bar A_t[\phi,S]=[\bar D_t,\bar\theta]-\delta_S
 \bar A_t\ ,
\ee
where we have introduced the bar notation to make it clear
that the functional dependence of $t$ is only coming through
the action. 

Globally, two kernels describe the same
RG flow between fixed points $i$ and  $f$ if their
corresponding connections $A_1$ and $A_2$ produce
$t$-ordered lines which are related by
\be
\label{finaleq}
\Phi_{fi}[A_1]= \Omega_f \Phi_{fi}[A_2]\Omega_i^{-1}\ ,
\ee
where $\Omega_{i,f}$ are $t$-independent field redefinitions of 
the fixed points actions at points $i$ and $f$. 
In perturbation theory, an operator basis is chosen and all field
redefinitions turn into finite coupling constant redefinitions. 
But this is nothing but a
change of perturbative scheme, that is of local counterterms,
which is also tantamount to a finite redefinition of coupling constants.

\bigskip

As a specific example, the infinitessimal field redefinition
\be
\label{sumif}
\theta^\alpha={1\over2}{\delta c(q^2)\over q^2}
{\delta S\over\delta\phi_{-q}}
-{\delta c(q^2)\over c(q^2)}\phi_q
\ee
corresponds to changing the cutoff function\cite{latmor}
$c\mapsto c+\delta c$.
(This may be derived from comparing (\ref{gas}) with some previous
observations,\cite{sumi,gol,bervil} or directly.) It is thus abundantly
clear that this change of scheme amounts to a finite
redefinition of coupling constants in the continuum theory,
with no physical effect. 

%\bigskip

Another useful version of the exact renormalization group may be obtained
by Legendre transformation of $S$.\cite{merg} The resulting action 
$\Gamma[\phi^c]$
has a simple interpretation as an infrared cutoff 
Legendre effective action, and has a corresponding 
flow equation.\cite{merg,bon} It is also the effective
average action.\cite{wet}
So far, all our analysis has been concerned with
the Polchinski form (\ref{polchinski}) 
and it would appear that the Legendre flow equation is left out of this
party. In general this is true because arbitrary transformations $\theta$
take us to exact renormalization group equations that no longer have
the specific form of (\ref{polchinski}), 
and thus the direct link to the Legendre effective action
is lost. However for the specific transformation (\ref{sumif}) the form
(\ref{polchinski}) is left alone: only the cutoff function changes. In this
case by utilising the 
Legendre transform relation,\cite{merg}
it should be possible to find directly corresponding transformations
for $\Gamma$ and $\phi^c$ which leave the
the Legendre flow equation invariant up to a change in
cutoff function $c\mapsto c+\delta c$.

\bigskip

Finally we would like to emphasise once again the 
freedom inherent to quantum field theories. Path integrals 
bring field redefinitions symmetries with  deep
consequences. The fact that exact renormalization group
equations follow from an instance of  Schwinger-Dyson type of
relations is striking. Their further structure and
transformation properties are fully determined by 
this underlying symmetry.

\section*{Acknowledgments}
J.I.L. acknowledges financial support from 
CICYT (contract AEN98-0431) and
CIRIT (contract 1998SGR-00026). T.R.M. acknowledges financial
support of PPARC grant GR/K55738. We both acknowledge
the financial support of a BC-MEC Acciones Integradas
grant MDR(A998/99)1799.

\end{document}